\documentstyle[prd,tighten,aps]{revtex}
\begin{document}
\draft

\twocolumn[\hsize\textwidth\columnwidth\hsize\csname
@twocolumnfalse\endcsname
\renewcommand{\theequation}{\thesection . \arabic{equation} }
\title{\bf Nonsingular Instantons for the Creation of Open Universes}

\author{Pedro F. Gonz\'alez-D\'{\i}az }
\address{Centro de F\'{\i}sica ``Miguel Catal\'an'',
Instituto de Matem\'aticas y F\'{\i}sica Fundamental,\\
Consejo Superior de Investigaciones Cient\'{\i}ficas,
Serrano 121, 28006 Madrid (SPAIN)}
\date{October 14, 1998}

\maketitle

\begin{abstract}
We show that the instability of the singular Vilenkin instanton
describing the creation of an open universe can be avoided using,
instead of a minimally coupled scalar field,
an axionic massless scalar field which gives rise to
the Giddings-Strominger instanton. However, if we replace
the singularity of the Hawking Turok instanton for an axionic
wormhole some interpretational and technical difficulties would
appear which can be overcome by introducing a positive
cosmological constant in the action.
This would make the instanton finite and
free from any instabilities.
\end{abstract}


\vskip2pc]

\renewcommand{\theequation}{\arabic{section}.\arabic{equation}}

\section{\bf Introduction}
\setcounter{equation}{0}

One of the most promising recent developments in theoretical
cosmology is the proposal by Hawking and Turok [1] of a singular
instanton for the creation of open universes. It has prompted
an almost inmediate influx of comments, criticisms, replies
and new work on the subject [2]. The relevance of this proposal
resides on the fact that it satisfies
the need for formulating open inflationary models
with $\Omega < 1$ which is required by recent observational
estimates of the present value of the Hubble constant [3-6]
and the density of gravitational lenses in
the universe [7,8]. Most available inflationary models
predicts a value for $\Omega$ very close to unity and cannot
therefore conform these observations.

Hawking and Turok used as solution to the Euclidean Einstein
equations [1]
\begin{equation}
ds^2=d\sigma^2+b(\sigma)^2\left(d\psi^2
+\sin^2 \psi d\Omega_2^2 \right),
\end{equation}
with $d\Omega_2^2$ the metric on the unit two-sphere. They
continued (1.1) into the metric for an open spatially
homogeneous universe with real scale factor by first matching
(1.1) to the inhomogeneous de Sitter-like Lorentzian solution
\begin{equation}
ds^2=d\sigma^2+b(\sigma)^2\left(-d\tau^2
+\cosh^2 \tau d\Omega_2^2 \right) ,
\end{equation}
accross the surface $\psi=\frac{\pi}{2}$, with the continuation
$\psi=\frac{\pi}{2}+i\tau$, and then continued (1.2) into the
metric of a Lorentzian isotropic open universe,
\begin{equation}
ds^2=-dt^2+a(t)^2\left(d\chi^2
+\sinh^2 \chi d\Omega_2^2 \right),
\end{equation}
with the rotation $\tau=\frac{\pi}{2}i+\chi$, $\sigma=it$,
$b(it)=ia(t)$, at the singular surface $\sigma=0$, where
$b=\sigma$.

The main criticism to this singular instantonic model has been
raised by Vilenkin [9] who showed that singular instantons of
the type suggested by Hawking and Turok but more manageable
and actually able to provide exact solutions as we shall see
in this paper, lead to the physically
unacceptable consequence that the resulting universe would be
overrun by a high density of expanding singular bubbles. In
order to avoid this problem, Garriga [10] regularized the
instanton singularity with matter, coupling a membrane to
the scalar field, and more recently [11], considering a
nonsingular five-dimensional instanton whose four-dimensional
sections are the Hawking Turok instanton. In this paper we
shall regularize the singularity of the latter
instanton by using quite
a different procedure: Instead of introducing any new
ingredient in the background theory, we will modify the nature
of the scalar field itself and/or its coupling to gravity. The
result is also a change in the instanton structure. We have
seen that, in order to avoid the singularity of the
Hawking Turok instanton or the
asymptotically flat instanton used by Vilenkin [9], one
should replace them for an axionic wormhole which becomes
a nonsingular finite instanton if a positive cosmological
constant is introduced in the action.

We outline the paper as follows. In Sec. II we review the
asymptotically flat instanton introduced by Vilenkin [9] by
using a closed form solution for it. Basing on the resulting
formalism we analyse in Sec. III the kind of change that we
must introduce in the matter field to obtain a nonsingular
instanton able to replace the Hawking Turok instanton, instead
of the Vilenkin solution.
What results from such a change is
the known Giddings-Strominger instanton [12], which we review
in terms of the conformal time, and discuss
in relation with the creation of an open universe along
the lines of the Hawking Turok model. This analysis is
extended in Sec. IV to include a positive cosmological
term in the distinct wormhole solutions.
Thus, two new candidates, the instantons first
considered by Halliwell and Laflamme [13] and by Myers [14],
emerge as suitable regularized substitutes of the singularity of
the Hawking Turok instanton. We also review these solutions in
terms of the conformal time and consider them in the framework
of the instantonic creation of open universes. In Sec. V we
consider the global structure of the introduced spacetimes
in the context of the Hawking Turok model.
We finally conclude in Sec. VI
where some comments on the obtained results are added.

\section{\bf The singular instanton}
\setcounter{equation}{0}

In this section we review the characteristics of the
asymptotically flat singular instanton introduced by
Vilenkin [9]. The analysis to follow is based on the
derivation of a closed form solution for such an instanton.
The model is very simple. It corresponds to a massless
scalar field $\phi$ which is minimally coupled to
Hilbert-Einstein gravity. The Euclidean action for this
model is [9]
\begin{equation}
S_E=\int d^4 x\sqrt{g}\left[-\frac{R}{16\pi G}
+\frac{1}{2}\left(\partial\phi\right)^2\right]
+S_{boundary}.
\end{equation}
The instantonic solution will be O(4) symmetric and, when
described by metric (1.1), the equations of motion for
the scale factor $b(\sigma)$ and massless scalar field
$\phi\equiv\phi(\sigma)$
are:
\begin{equation}
\phi''+3\frac{b'}{b}\phi'=0
\end{equation}
\begin{equation}
b''=-\frac{8\pi G}{3}b\phi'^{2},
\end{equation}
with the primes standing for derivatives with respect to
the Friedmann-Robertson-Walker time $\sigma$. From (2.2)
we have $\phi'=-\frac{C}{b^3}$, with $C$ a real integration
constant. After integration, equation (2.3) then becomes:
\begin{equation}
b'^{2}-\frac{l_p^2 C^2}{b^4}=1,
\end{equation}
where $l_p=\sqrt{4\pi G/3}$ and the integration constant
has been chosen to be unity in order to satisfy the
necessary conditions that correspond to an asymptotically
flat solution; i.e.: $b(\sigma)\approx\sigma$ and
$\phi(\sigma)\rightarrow 0$ as $\sigma\rightarrow\infty$.

We now note that the equation of motion (2.4) admits an
exact analytical solution. This can be most easily seen
if we introduce the conformal time $\Sigma$, defined by
\begin{equation}
\Sigma=\int_{0}^{\sigma}\frac{d\sigma '}{b(\sigma ')},
\end{equation}
so that
\begin{equation}
b^2\dot{b}^2-b^4 -l_p^2 C^2=0,
\end{equation}
with $^{.}=d/d\Sigma$. The solution to (2.6)
reads:
\begin{equation}
b(\Sigma)=\sqrt{Cl_p}\sinh^{\frac{1}{2}}(2\Sigma).
\end{equation}

From (2.5) we also obtain
\[\sigma-\sigma_{*}=\]
\begin{equation}
\sqrt{Cl_p}\left\{\frac{1}{2}\left[F(\alpha,r)
-2E(\alpha,r)\right]+\frac{b(\Sigma)\sqrt{C^2 l_p^2+b(\Sigma)^4}}
{\sqrt{Cl_p}(Cl_p+b(\Sigma)^2)}\right\},
\end{equation}
where $F$ and $E$ are the elliptic integral of the first and
second kind, respectively,
$\alpha=\arccos\left[\frac{Cl_p-b(\Sigma)^2}{Cl_p+b(\Sigma)^2}\right]$,
$r=\sqrt{2}/2$, and $\sigma_{*}$ is an integration constant.
Clearly, for $\Sigma=0$, $\sigma=\sigma_{*}$, while for
$\Sigma\rightarrow\infty$, $\sigma\rightarrow\infty$.

An analytical solution for the field $\phi$ can also be
obtained in terms of the conformal time $\Sigma$. From
the conservation of the matter field charge,
$b^2\dot{\phi}=-C$, one gets
\begin{equation}
\phi=-\frac{1}{4l_p}\ln\left(\frac{\cosh 2\Sigma-1}
{\cosh 2\Sigma+1}\right)+\phi_{*}.
\end{equation}
In order to satisfy the boundary conditions in the
asymptotic region one should set $\phi_{*}=0$. Solutions
(2.7) and (2.9) satisfy then all the wanted requirements;
i.e.: as one moves towards smaller values of $\sigma$,
$b(\sigma)$ decreases and $\phi(\sigma)$ grows until $b$
vanishes and $\phi$ logarithmically diverges at
$\sigma=\sigma_{*}$.

As it was already noted in Ref. [9], the only contribution
to the Euclidean action comes from the boundary term at
$\sigma=\sigma_{*}$, and is given by
\begin{equation}
S_E=S_{boundary}=
-\frac{1}{8\pi G}\partial_{normal}V_B,
\end{equation}
(with $V_B$ the volume of the boundary)
which, using (2.5), re-produces the value
$S_E=\sqrt{3\pi^2/4G}C$ obtained by Vilenkin [9].

Of course, this instanton is an O(4)-instanton which
can be described by the Hawking Turok metric
\begin{equation}
ds^2=b(\Sigma)^2\left(d\Sigma^2+d\psi^2
+\sin^2 \psi d\Omega_2^2\right) ,
\end{equation}
which should first be matched to the spatially inhomogeneous
de Sitter-like solution
\begin{equation}
ds^2=b(\Sigma)^2\left(d\Sigma^2-d\tau^2
+\cosh^2 \tau d\Omega_2^2\right) ,
\end{equation}
accross the surface $\psi=\frac{\pi}{2}$, by making the
continuation $\psi=\frac{\pi}{2}+i\tau$, and then continued
again into the metric of a Lorentzian isotropic open universe
\begin{equation}
ds^2=a(\Sigma)^2\left(-d\eta^2+d\chi^2
+\sinh^2 \chi d\Omega_2^2\right) ,
\end{equation}
where $d\eta=\frac{d\sigma}{a(\Sigma)}$, by using the additional
continuation $\Sigma\rightarrow -\Sigma$ (i.e. $\sigma=it$),
$\tau=\frac{\pi}{2}i+\chi$, and $b(-\Sigma)=ia(\Sigma)$, at the
singular surface $\sigma=\sigma_{*}$.

The singular behaviour of this instanton is exactly the same
as that of the Hawking Turok instanton, but has the physically
unacceptable property of predicting the creation of an open
universe filled with expanding singular bubbles [9].

\section{\bf Replacing the singularity with a wormhole}
\setcounter{equation}{0}

From the conservation of the
matter field charge $\frac{d}{d\Sigma}(b^2\dot{\phi})=0$, one
can generally write: $b^2\dot{\phi}=\mp C$, with $C$ an
integration constant. For ordinary fields where $C$ is real,
we obtain the singular instanton reviewed in Sec. II. However, if
the considered instanton arises from an axionic conserved
charge, to examine instantonic transitions with fixed axion charge,
we must then modify the Euclidean action (2.1) by an additional
surface term which is given by
\begin{equation}
S_A=\left.-b^2\phi\dot{\phi}\right|_{0}^{T},
\end{equation}
and hence [14] one obtains from the conservation of the momentum
conjugate to $\phi$ that we must choose the constant $C$ to be
pure imaginary, i.e.: $b^2\dot{\phi}=\pm iC$, and therefore,
instead of (2.6), we get for the equation of motion for the
scale factor
\begin{equation}
b^2\dot{b}^2-b^4+l_p^2C^2=0,
\end{equation}
and hence, instead of the singular instanton solution (2.7),
(2.9), we have as a closed form solution for the instanton:
\begin{equation}
b(\Sigma)=\sqrt{Cl_p}\cosh^{\frac{1}{2}}(2\Sigma)
\end{equation}
\begin{equation}
\phi(\Sigma)=\frac{i}{2l_p}\arcsin(\tanh 2\Sigma)+\phi_{*} ,
\end{equation}
where in this case $\phi_{*}=-\frac{i\pi}{4l_p}$, and
\[\sigma-\sigma_{*}=\]
\begin{equation}
\sqrt{Cl_p}\left\{\frac{1}{\sqrt{2}}\left[F(\alpha,r)
-2E(\alpha,r)\right]+\frac{\sqrt{Cl_p}\sinh 2\Sigma}
{b(\Sigma)}\right\},
\end{equation}
with $\alpha=\arcsin\left(\frac{b(\Sigma)^2-Cl_p}{b(\Sigma)^2}\right)$
and $r=\frac{\sqrt{2}}{2}$. Again, for $\Sigma=0$, $\sigma=\sigma_{*}$,
and for $\Sigma\rightarrow\infty$, $\sigma\rightarrow\infty$. However,
the imaginary field $\phi$ will now become $\phi_{*}$
at $\Sigma=0$ and vanish asymptotically. The differences between
the instantons corresponding to the real scalar field and
those associated with pure-imaginary values of the scalar
field have been discussed in [15].
The scale factor (3.3) corresponds to a nonsingular instanton which
is nothing but the known Giddings-Strominger wormhole [12]. Using (2.10)
for this instanton, we obtain $S_{boundary}=0$,
which, by taking into account (3.1),
in turns implies that the Euclidean action
$S_E=\sqrt{\frac{3\pi}{16G}}C$. Thus,
the nonsingular instanton (3.3), (3.4) appears to be more stable than
the singular instanton (2.7), (2.9) and, moreover, would be free from
the instabilities Vilenkin pointed out [9] for this.

The new nonsingular instanton will also be able to describe the
creation of an open inflationary universe. In fact, for solutions
(3.3) and (3.4) one can also continue the metric for the spatially
inhomogeneous solution into the metric of a nonsingular
Lorentzian isotropic open universe
\begin{equation}
ds^2=a(\Sigma)^2\left(-d\Sigma^2+d\chi^2
+\sinh^2\chi d\Omega_2^2\right)
\end{equation}
by continuing (2.11) so that $C\rightarrow -C$, $\tau=\frac{\pi}{2}i
+\chi$, $b(\Sigma,C)=ia(\Sigma,-C)$, at the minimal nonsingular
surface $\sigma=\sigma_{*}$. Thus, if we replace the singularity
of the Hawking Turok instanton for one these wormholes, then
the inestability pointed out by Vilenkin would be avoided.
This solution would somewhat be related to that recently
suggested by Garriga [10] in terms of a scalar field coupled
to a membrane in that both solutions avoid the instability problem
by smoothing the singularity out.

Thus, when the Giddings-Strominger wormhole with scale factor
(3.3) is used in the Hawking Turok metric (2.11), it, too, is
able to describe creation of an open inflationary universe.
This is a novel result which does not contradict the original
interpretation given by Giddings and Strominger [12] for
Euclidean wormholes, according to which a wormhole instanton
describes a microscopic connection between two asymptotically
flat regions belonging to the same universe, or to two
different universes, but can be thought to complete it
whenever one allows the wormhole to also become the seed
for the creation of an open universe along the lines
discussed above and, in more detail, in Sec. V. What is
actually new here is the use being given to a
wormhole when this is continued from the
Euclidean region into a spacetime described by metric
(3.6), even if we keep the original interpretation for
wormholes given by Giddings and Strominger.

Replacing the singularity of the Hawking Turok instanton for
an axionic wormhole leads in turn to other difficulties. First
of all, we have an interpretational problem. Since the
wormhole is supposed to connect two nearly flat spacetime
regions, one would consider what is this wormhole connected
to. This problem could be solved by simply connecting the
wormhole to another copy of the instanton; this possibility
was pointed out to us by Vilenkin [16], but one does not
know for sure how it could be implemented physically. On the other
hand, one might alternatively adopt another approach to the
problem by renouncing the idea that the universe is created
from nothing, replacing it for the notion of universe creation
from a pre-existing large universe which the wormhole would
be connected to.

Another more serious difficulty with the idea of an
axionic wormhole smoothing out
the singularity is that the Giddings-Strominger instanton
tuns out to be instable against fluctuations of the metric.
Such a wormhole possesses a negative mode which would
induce instabilities of flat space to the creation of baby
universes [17].

\section{\bf Nonsingular finite instantons}
\setcounter{equation}{0}

The wormhole instanton discussed in Sec. III corresponds to
a solution of the Euclidean Einstein equations with
vanishing cosmological constant and no scalar field
potential, $V(\phi)=0$. The asymptotically flat or
anti-de Sitter behaviour [18] of wormhole spacetimes is
lost when a nonzero potential, $V(\phi)>0$, and/or a positive
cosmological constant, $\Lambda >0$, is included in the
corresponding action. The resulting instantons are then
seen to be nonsingular and connect given surfaces of finite
size. If we take them to replace the singularity of the
Hawking Turok instanton, such instantons would not show any
of the problems pointed out at the end of the precedent
section. Taking advantage of this fact, we shall include
a positive cosmological term in the wormhole instantonic
action and replace the singularity of the Hawking Turok
instanton for half of the resulting nonsingular finite
instantons. We will divide the instantons in two equal
halves by slacing the space through the minimal surface
at the neck, i.e. at the surface of minimal size.

There are two possible such instantons. We first consider
the case where a massless scalar field is conformally
coupled to Hilbert-Einstein gravity, adding a positive
cosmological term. The Euclidean action is then
\[S_E=-\frac{1}{16\pi G}\int d^4 x\sqrt{g}
\left[(1-l_p^2\phi^2)R-2\Lambda\right]\]
\begin{equation}
+\frac{1}{2}\int d^4 \sqrt{g}(\nabla\phi)^2
-\frac{1}{8\pi G}\int d^3 x\sqrt{h}(1-l_p^2\phi^2)TrK ,
\end{equation}
with $K$ the second fundamental form.

For the metric (1.1) one obtains then the Halliwell-Laflamme
nonsingular instanton [13]. We have found the closed form
solution in terms of the conformal time $\Sigma$ for this
instanton
\begin{equation}
b(\Sigma)=\frac{\xi+1}{\sqrt{2\lambda/3}}
\left[1+\xi(1-2sn^2\eta)\right]^{-\frac{1}{2}} ,
\end{equation}
where $sn$ is the Jacobean elliptic function, and we have
introduced the following definitions:
\begin{equation}
\xi=\sqrt{1-4R_0\lambda}
\end{equation}
\begin{equation}
\eta=\sqrt{\frac{\xi+1}{2}}\Sigma,
\end{equation}
with $R_0$ an integration constant and $\lambda=l_p^2\Lambda/6\pi^2$.
For solution (4.2) $K$ and $\lambda$ are subject to the
condition $4K\lambda < 1$. Thus, $b$ will vary between
the two extreme values $b(0)=\frac{\xi+1}{\sqrt{2\lambda(\xi+1)/3}}$
(minimal surface) and $b(K)=\frac{\xi+1}{\sqrt{2\lambda(1-\xi)/3}}$
(maximal surface), the value $\eta=K$ being the quarter-period
of the Jacobean elliptic function $sn\eta$.
This instanton can also be used as the solution for the
scale factor in the Hawking Turok metric (2.11) in order
to describe creation of open inflationary universes.
Metric (2.11) can then be continued first into (2.12) with
the rotation $\psi=\frac{\pi}{2}+\tau$, without changing the
scale factor (4.2), and then into (2.13) by using the
transformation of the parameters entering the
Halliwell-Laflamme instanton (4.2)
$\Sigma\rightarrow\Sigma$, $\xi\rightarrow\xi$,
$\Lambda\rightarrow -\Lambda$, $K\rightarrow -K$, which
correspond to the time rotation $\sigma\rightarrow it$
and
\begin{equation}
b(\Sigma,-\Lambda,-K)=ia(\Sigma,\Lambda,K)
=ib(\Sigma,\Lambda,K).
\end{equation}

As it was already pointed out by Halliwell and Laflamme [13],
this instanton has the problem that the effective gravitational
constant can become negative when the field $\phi$ exceeds
some fixed value. This problem can be avoided if, instead of
a conformally coupled scalar field, we use a minimally coupled
axionic field with a positive cosmological constant. The resulting
instantons were first considered by Myers [14], though he
did not get the solutions in closed form.

In order to check that the axionic finite instantons can be
adapted to the Hawking Turok continuation leading to metric
(2.12) for a Lorentzian isotropic open universe, let us
obtain a closed form solution for their simplest
three-dimensional case whose Euclidean action can be written as
\[S_E=-\frac{1}{16\pi G}\int d^3 x\sqrt{g}
(R-2\Lambda)\]
\begin{equation}
-\frac{1}{2}\int d^3 x\sqrt{g}(\nabla\phi)^2
-\frac{1}{8\pi G}\int d^2 x\sqrt{h}K+S_A .
\end{equation}
where $S_A$ is an axionic boundary term which, for the
three-dimensional instantonic metric
\begin{equation}
ds^2=b(\Sigma)^2\left(d\Sigma^2+d\theta^2
+\sin^2\theta d\varphi^2\right) ,
\end{equation}
can be written as $S_A=-b\dot{\phi}\phi |_{0}^{T}$, in the
gauge where the lapse function is set to $N=1$. Restricting
the Myers instanton to be three-dimensional is here
required by the fact that three is the highest possible
spacetime dimensionality for this instanton which is
able to allow exact solutions of the field equations.
However, all the geometric properties of the
three-dimensional Myers instanton, particularly its
capability to describe creation of open inflationary
universes, can be readily extrapolated to the
physically more interesting four-dimensional case.

From the equations of motion for $\phi$ and $b$, we get
\begin{equation}
\dot{\phi}=\frac{im}{b^2}
\end{equation}
\begin{equation}
\dot{b}^2+m^2+\Lambda b^4=1 .
\end{equation}
The solution to (4.8) and (4.9) can still be given in closed
form as:
\begin{equation}
b(\Sigma)=\frac{1}{\sqrt{2\Lambda}}\sqrt{1+\zeta(1-2sn^2\delta)}
\end{equation}
\begin{equation}
\phi=i\arctan\left[\frac{\tan\left(\sqrt{\Lambda}\sigma\right)+\zeta}
{2m\sqrt{\Lambda}}\right]+\phi_{*} ,
\end{equation}
where for the relation between $\sigma$ and $\Sigma$ we have
\begin{equation}
\sigma-\sigma_{*}=-\frac{1}{2\sqrt{\Lambda}}
\arcsin\left(\frac{1-2\Lambda b^2}{\zeta}\right) ,
\end{equation}
and we have introduced the parameters:
\begin{equation}
\zeta=\sqrt{1-4m^2\Lambda}
\end{equation}
\begin{equation}
\delta=\sqrt{\frac{1+\zeta}{2}}\Sigma .
\end{equation}

We can see from (4.10) that the scale factor oscillates now
between the extreme values $b(0)=\sqrt{\frac{1+\zeta}{2\Lambda}}$
(maximal surface) and $b(K)=\sqrt{\frac{1-\zeta}{2\Lambda}}$
(minimal surface), the value $\delta=K$ again being the
quarter-period of the Jacobean elliptic function $sn\delta$.
The Euclidean action for this instanton can be easily
computed [14]. It is given by
\[S_E=\frac{\pi}{2G}\left(m-\frac{1}{2\sqrt{\Lambda}}\right),\]
which is finite and negative. Thus, this instanton will contribute
the Euclidean path integral for the creation of open universes.

If we now match the Euclidean metric (4.7) to the
three-dimensional spatially inhomogeneous solution
\begin{equation}
ds^2=b(\Sigma)\left(d\Sigma^2
-d\tau^2+\cosh^2\tau d\varphi^2\right) ,
\end{equation}
accross the surface $\theta=\frac{\pi}{2}$, by making the
continuation $\theta=\frac{\pi}{2}+i\tau$, then, given
the form of the scale factor (4.10), the resulting
metric (4.15) can be continued into the metric of a
Lorentzian isotropic open universe in three dimensions:
\begin{equation}
ds^2=a(\Sigma)\left(-d\Sigma^2
+d\chi^2+\sinh^2\chi d\varphi^2\right) ,
\end{equation}
by using the additional continuation $\sigma\rightarrow it$,
$\Lambda\rightarrow -\Lambda$, $m\rightarrow im$,
$\tau\rightarrow\frac{\pi}{2}i+\chi$, so that
$\Sigma\rightarrow\Sigma$ and
\begin{equation}
b(\Sigma,\Lambda,m)\rightarrow ia(\Sigma,-\Lambda,im)
=ib(\Sigma,\Lambda,m) .
\end{equation}

Thus, we obtain the wanted Lorentzian isotropic
metric for a three-dimensional open universe, provided
we take a negative cosmological constant and an ordinary
massless nonaxionic field. We note, moreover, that starting
with action (4.6), which we perturb introducing O(3)-symmetric
fluctuations about the classical instanton solution, $b=b_{cl}+r$,
and evaluate in the quadratic part, one can follow the
semiclassical Rubakov-Shvedov procedure [17] to finally obtain
a frequency squared for the fluctuation mode given by
\begin{equation}
\omega^2\propto -\Lambda\left(1
+\frac{m^2}{\Lambda b_{cl}^4}\right).
\end{equation}
So, even at the maximal surface of the instanton, $\omega^2$
will always be definite positive when we simultaneously rotate
$\Lambda\rightarrow -\Lambda$ and $m\rightarrow im$. Therefore,
the nucleation of this instanton will not induce any
instabilities in the resulting Lorentzian open universe.

Although we have not succeeded in obtaining a closed
form solution for the corresponding four-dimensional case,
it is not difficult to convince oneself that the
four-dimensional Myers instanton shares the same general behaviour
as that we have discussed for (4.10), (4.11) and, therefore, it
can be used as a suitable substitute for the singularity of
the Hawking Turok instanton for the creation of an open
universe from nothing.

\section{\bf Global structure of wormhole-Hawking-Turok
spacetimes}
\setcounter{equation}{0}

We have seen that asymptotically flat Euclidean wormholes
and the instantonic tunnelings that result from introducing a
positive cosmological constant in the wormhole action can
all be used to describe the evolution of the scale factor
that enters the Hawking Turok metric, and hence are able
to represent creation of an open inflationary universe by means
of a regular instanton. In what follows, we shall discuss
the geometrical significance of this result.

First of all, since in the coordinates $\sigma$,$\tau$ of
metric (1.2) the surface $\sigma=0$ is a null surface,
one can re-define the second continuation leading to
metric (1.3) by extending the coordinates beyond the
surface $\sigma=0$ which is singular in the Hawking
Turok geometry. Then, by using the conformal time $\Sigma$,
defined in (2.5), one can introduce Kruskal-like coordinates [2]
\begin{equation}
U=e^{-\tau+\Sigma} ,\;\;\; V=-e^{\tau+\Sigma} ,
\end{equation}
so that metric (1.2) and its conformal-time equivalent
(2.12) can both be re-written as
\begin{equation}
ds^2=b^2 e^{-2\Sigma}\left(-dUdV+
\frac{1}{4}(U-V)^2 d\Omega_2^2\right),
\end{equation}
where
\begin{equation}
e^{-2\Sigma}=-\frac{1}{UV} .
\end{equation}

In order for continuing (5.2) into the metric of an open
spatially homogeneous universe with real, positive definite
scale factor and real coordinates for the instantons considered
in the present paper and also the Hawking Turok instanton,
we first introduce the new coordinates $T$,$\chi$ [2] through
\begin{equation}
U=e^{T-\chi} ,\;\;\; V=e^{T+\chi} ,
\end{equation}
and then rotate $\sigma$ and $b$ so that $\sigma=it$,
$b(it)=ia(t)$, while keeping coordinate $\tau$ real and
unchanged. Hence we get the required metric for an open
spatially homogeneous universe:
\begin{equation}
ds^2=a^2\left(-dT^2+d\chi^2+\sinh^2\chi d\Omega_2^2\right).
\end{equation}

The global structure of our wormhole-derived spacetimes
can now be investigated by considering the $U$,$V$
diagram given in Fig. 1. It corresponds to the simplest
$V(\phi)=0$ version of a Hawking Turok instanton in which
the singularity has been replaced for either the throat of
an asymptotically flat wormhole, or for an extremal surface
of the finite nonsingular instantons dealt with in Sec. IV.
In the sector defined by coordinates $T$,$\chi$ on such a
diagram the dotted line at $U=V$ will correspond to $\chi=0$,
that is the origin of the spherically symmetric $\chi$,$\theta$,
$\varphi$ coordinate system. For all the cases being considered
the dashed line at $UV=-1$, that is at $\sigma=\sigma_{*}$ on
the $\tau$,$\Sigma$ region, which is separated from the region
described by coordinates $T$,$\chi$ along the axis $V=0$, is
either the throat of the wormhole (that is to say, the
maximal surface of the corresponding baby universe in
Lorentzian time), or one of the extremal surfaces (maximal
for the axionic instanton and minimal for the Halliwell-Laflamme
instanton) of the considered two finite instantons. On that
line the field $\phi$ always takes on finite values, both
for the axionic and the conformally-coupled field cases.
This is the crucial point that distinguishes the spacetime
structure of the instantons represented in Fig. 1 from
that of the Hawking Turok and Vilenkin instantons for
which $\phi$ would diverge on the surface $S_0$ (Fig. 1).

Since near the maximal surface of the finite nonsingular
instantons the cosmological constant dominates the
curvature, that surface corresponds to a de Sitter space
deformed by the residual effect of the axion or conformal
field, so allowing the resulting instanton to have O(4)
symmetry. Near the minimal surface, it is respectively
the axion and the conformal field which dominates the
curvature, making a deformed baby universe. It follows
that what essentially distinguishes the instantons
considered in this paper from the Vilenkin and Hawking
Turok instantons is the presence in the former instantons
of a baby universe which (see Fig. 1) (a) replaces the deformed de
Sitter space of the Hawking Turok solution (for $\Lambda=0$),
(b) intercalates after this space (for $\Lambda>0$ and
conformally coupled field), and (c) is an initial
boundary from which the system tunnels into the deformed
de Sitter space (for $\Lambda>0$ and axionic field).

The region $\tau$,$\Sigma$ can be connected to the
Euclidean sector on the surface defined by $\tau=0$,
$\phi=\phi_{*}$, $U=-V$. The wormhole (case (a))
develops from that surface approaching an asymptotically
flat region where $\phi=0$, and the finite instantons
do approaching either the maximal finite surface (case
(b)) or the minimal surface (case (c)). In the latter
two cases there exists the interesting possibility of
sewing together two copies of the whole solution
illustrated in Fig. 1 on the Euclidean sector, to give
rise to an overall process which can be interpreted as
the creation of two open inflationary universes from
either tunneling between two baby universes through
a deformed de Sitter universe (case (b)) or tunneling
between two deformed de Sitter universes through a
baby universe when the Myers instanton is considered.

On the other hand, we always can have a four-geometry
on the Lorentzian sector which can be regarded as a
valid interpolating geometry (see Fig. 1) between
one of the spatial hypersurfaces in the $T$,$\chi$
region and either the wormhole throat, or the maximal
(axionic field) or minimal (conformal field) surface
at $\tau<0$. Unlike in the original Hawking Turok
spacetime, in any of the considered cases, any of the
three-geometries on $T$,$\chi$, together with the
interpolating four-geometry and the three-geometry at
$\Sigma=0$, is a regular geometry. Note that if the
surfaces at $\Sigma=0$ would be replaced by a singularity,
then this singularity would engulf the whole spacetime,
just as the singularity of the Vilenkin model does [9].

In the language of the prescription for the wave function
of the universe, the interpolating four-geometry may be
used to calculate the probability for the propagation
from, respectively, the wormhole throat or the extremal
surfaces of finite instantons to the given three-geometry
on the region with $T$,$\xi$ coordinates. Such a calculation
would be performed by looking at the imaginary part of the
corresponding action.

Finally, let us briefly comment on the four-dimensional
axionic finite instanton within the context of the Hawking
Turok procedure, since this instanton appears to be of most
physical interest. Although the four-dimensional Myers
solution is not exactly solvable in terms of the scale
factor in a Robertson-Walker metric, according to Myers [14]
one can still write the spherically symmetric metric for
such a solution in the form:
\[ds^2=\left(1-\frac{\Lambda r^2}{3}-
\frac{m^2}{r^4}\right)^{-1}dr^2+r^2 d\Omega_3^2 ,\]
where $d\Omega_3^2$ is the metric on the unit three-sphere
and $m$ is an integration constant. Extreme surfaces occur
at $g_{rr}=\infty$, which correspond to coordinate
singularities as the components of the curvature tensor
become all finite there. Then, similarly to as it happens
in the three-dimensional case, the radial coordinate, and
actually the scale factor, must vary between the extreme
values defined by the apparent singularities at $g_{rr}^{-1}=0$.
Specialising e.g. to the particular, simplest case where
$\frac{3}{4}m^2\Lambda^2=1$, the extreme values become
\begin{equation}
r_{\pm}=\Lambda^{-1}\pm\left(6m^3 /\Lambda\right)^{\frac{1}{3}}.
\end{equation}

Moreover, a scale factor with extreme values as given by (5.6)
(or by more complicate expressions for more general cases)
can also be used in the procedure leading from (1.1) to (1.3).
Thus, the four-dimensional Myers instanton can
be considered as a rather natural ingredient of the global
spacetime structure (for Euclidean sector (c) of Fig. 1)
of the nonsingular solutions described in this Section, and be
therefore used as a convenient substitute for the singularity
of the Hawking Turok instanton.

\section{\bf Conclusions}
\setcounter{equation}{0}

Wormholes are genuine components of the quantum spacetime
foam. So far, they have been used for a variety of purposes,
including: as spacetime channels to transfer the information
fell in a black hole into a different universe, to induce
loss of quantum coherence in ordinary matter at low energy,
or most popularly, to implement a mechanism for determining
the coupling constants [12,19] and the
vanishing of the cosmological constant [20]. In this paper
wormholes and their finite instantonic generalizations with
a positive cosmological constant are added the new use of playing
the role of the instantons required by the mechanism suggested
by Hawking and Turok for the creation of open universes [1].
The proposal is based on the realization that the singular,
asymptotically flat instanton discovered by Vilenkin [9] is
nothing but the ordinary scalar field counter-part of the
axionic Giddings-Strominger instanton [12] that corresponds
to a different formalation of the problem.

However, if we choose the nonsingular axionic wormhole to
replace the singularity of the Hawking Turok instanton, new
difficulties arise which are of both interpretational and
technical nature. In order to overcome these difficulties
one should introduce either a nonzero generic potential
for the scalar field or a nonvanishing positive
cosmological constant, both making the instanton finite.
We have considered the solutions that result in the latter
case for scalar fields which are both conformally [13] and
minimally [14] coupled to gravity, suggesting these solutions
as suitable candidates to replace the singularity of the
Hawking Turok instanton for the creation of an open universe.
When included in the continuation procedure that gives rise
to open universes, the instantons arising from minimally
coupling an axionic massless scalar field to gravity in
the presence of a positive cosmological constant appear
to be free from any of the shortcomings that could lead
to interpretational or technical difficulties, within
the considered model for the creation of open universes.
It appears therefore that the most likely process for
the creation of single or paired open universes would
include tunneling from the baby universes which are thought [21]
to pervade the quantum structure of the vacuum spacetime
foam.

Our conclusions are based on the analysis of the different
solutions in closed form. However, in order for the
creation model to generate an inflationary process and
an acceptable value of the cosmological parameter
$\Omega < 1$ as well, one ought to consider the path
integral for Einstein gravity coupled to a scalar field
with a nonvanishing matter potential whose shape will
ensure O(4) invariance, which does not allow having
classical solutions in closed form. Therefore, acceptable
predictions for $\Omega$ in the framework of an
appropriate open inflationary model could only be
achieved if we include a simple generic potential $V(\phi)$
with just a true minimum at $V=0$ in our wormhole-like
model, analysing all allowed predictions in accordance
with procedures similar to those used by Hawking and Turok
[1,2].

\acknowledgements

\noindent For useful correspondence the author thanks
A. Vilenkin. This reasearch was supported by DCICYT under
Research Project No. PB94-0107.

\vspace{2cm}

\begin{center}

{\bf Legend for Figure}

\end{center}

\noindent Fig. 1. Global structure for the nonsingular instantons.
The Euclidean region continues into the Lorentzian one on the
surface $U=-V$, $\tau=0$, $\phi=\phi_{*}$. It corresponds to:
(a) an asymptotically flat wormhole instanton, (b) a finite
Halliwell-Laflamme instanton, and (c) a finite Myers instanton.

\end{document}